\documentclass[12pt]{article}
\usepackage[english]{babel}
\usepackage[latin1]{inputenc}
\usepackage{graphicx}
\usepackage{rotating}
\usepackage{pdflscape}
\usepackage{graphics}
\usepackage{setspace}
\usepackage{caption}
\usepackage{array}
\newcolumntype{L}[1]{>{\raggedright\let\newline\\\arraybackslash\hspace{0pt}}m{#1}}
\newcolumntype{C}[1]{>{\centering\let\newline\\\arraybackslash\hspace{0pt}}m{#1}}
\newcolumntype{R}[1]{>{\raggedleft\let\newline\\\arraybackslash\hspace{0pt}}m{#1}}

\newcommand{\adev}{$r_i^{*dev}$}
\newcommand{\apea}{$r_i^{*pea}$}
\newcommand{\aasc}{$r_i^{*ans}$}
\newcommand{\awil}{$r_i^{*wil}$}

\newcommand{\rqu}{$r_i^{**q}$}

\RequirePackage[OT1]{fontenc}
\RequirePackage{amsthm,amsmath,amsfonts,amssymb}
\RequirePackage[authoryear]{natbib}
\RequirePackage[colorlinks,citecolor=blue,urlcolor=blue]{hyperref}
\usepackage{bm}

\numberwithin{equation}{section}
\theoremstyle{plain}
\newtheorem{thm}{Theorem}[section]
\newtheorem{definition}{Definition}[section]

\begin{document}


\title{A residual for outlier identification in zero adjusted regression models}






\author{Gustavo H. A. Pereira$^1$ \and Juliana S. Rodrigues$^{1,2}$ \and  Manoel Santos Neto$^{1,3}$ \and Denise A. Botter$^4$ \and Mônica C. Sandoval$^4$}
\date{}

\maketitle

\vspace{-11.5mm} \noindent 
\begin{center}
$^1$Department of Statistics, Federal University of São Carlos \\
$^2$Department of Applied Mathematics and Statistics, University of São Paulo \\
$^3$Department of Statistics, Federal University of Campina Grande \\
$^4$Department of Statistics, University of São Paulo \\
\end{center}
\vspace{1mm}

\begin{abstract}
Zero adjusted regression models are used to fit variables that are discrete at zero and continuous at some interval of the positive real numbers. Diagnostic analysis in these models is usually performed using the randomized quantile residual, which is useful for checking the overall adequacy of a zero adjusted regression model. However, it may fail to identify some outliers. In this work, we introduce a residual for outlier identification in zero adjusted regression models. Monte Carlo simulation studies and an application suggest that the residual introduced here has good properties and detects outliers that are not identified by the randomized quantile residual. 
\end{abstract}

\vspace{0mm} \noindent {\textbf{Key words}}: diagnostic analysis, outliers, randomized quantile residual, zero adjusted regression models.

\vspace{3mm}

\section{Introduction}
\label{sec:int}

Zero adjusted regression models are used to fit variables that are discrete at zero and continuous at some interval of the positive real numbers. They are useful in many areas such as insurance \citep{bortoluzzo2011estimating}, microbiology \citep{rocha2017association}, credit risk \citep{tong2016exposure}, hydrology \citep{serinaldi2014simulating}, biodiversity \citep{rubec2016using} and geology \citep{decarlo2015coral}. 
When the support of the response variable is the interval $[0;\infty[$, some possible models that can be used are the zero adjusted gamma regression model \citep{Tong:2013aa}, the zero adjusted gaussian inverse regression model \citep{Hinde2006aa} and the zero adjusted Birnbaum-Saunders regression model \citep{tomazella2018zero}. The zero adjusted beta regression model \citep{cook2008regression,Ospina:2012aa}, also known as the zero inflated beta regression model, is usually used when the response variable assumes values in the interval $[0;1[$.

Residuals play an important role in checking model adequacy and in the identification of outliers and influential observations. For each observation of the sample data, a residual is a measure of disagreement between the observed and the fitted value by a regression model \citep{Cook:1982ab}.
When the distribution of the residual is well approximated by a known probability distribution, we can classify an observation as a possible outlier if the absolute value of its residual exceeds a threshold. Otherwise, an observation is classified as a possible outlier if the absolute value of its residual is considerably greater than that of almost all observations. Since outliers are observations poorly fitted by the model, it is important to detect them for many reasons. First, outliers can be a result of an operational error that should be solved before using data and the regression model to make inference. Second, especially if many outliers are identified, they can suggest that the fitted regression model is not adequate for these data. Finally, outliers are usually also influential, which are observations that cause major changes in inferential analysis results \citep{kutner2004applied}.

In zero adjusted regression models, there are two fitted values for each observation that are the main interest. The first is the estimate of the probability of the observation assuming the zero value. The other is the estimate of the expected value of the response variable given that the observation assumed a non-zero value. For observations that assumed the zero value, only the former is useful, but both are important for the other observations. As the number of fitted values of interest is different according to the assumed value, it is difficult to define a single residual for all observations.

The randomized quantile residual  \citep[][]{Dunn:1996aa} is useful for checking the overall adequacy of a zero adjusted regression model \citep{Ospina:2012aa} and can be obtained for all observations. However, it may fail to identify some outliers when the probability of the response variable assuming positive values close to zero is very small (see Section \ref{sec:2}). Therefore, for outlier identification, it is convenient to define a different residual for the discrete and for the continuous component of the model. For the discrete component, a residual defined for regression models for binary dependent variables~\cite[][Chapter 5]{Hosmer_2000} can be used. Similarly, for the continuous component, we can use a residual commonly used in regression models for continuous response variables \citep{McCullagh:1989aa}. However, for observations that assumed a non-zero value, this strategy is not appropriate because the residual is not a function of the estimate of the probability of the observation assuming the zero value.

For the zero adjusted beta regression model, \citet{Ospina:2012aa} proposed a residual that is a function of a residual proposed by \citet{esp+fer+cri08a} for beta regression and the estimate of the probability of the observation assuming the zero value. Nonetheless, simulation studies (see Section \ref{sec:3}) suggest that this residual is not good for outlier identification in the zero adjusted beta regression model.

In this work, we introduce a class of residuals for outlier identification in zero adjusted regression models. This new class is a function of a residual defined for the continuous component of the model and the maximum likelihood estimate of the probability of the observation assuming the zero value. Simulation studies  (Section \ref{sec:3}) suggest that one of the residuals of this class is adequate for outlier identification in this class of regression models. Moreover, this residual detects outliers that are not identified by the randomized quantile residual (see Section \ref{sec:4}).

The remainder of this paper is organized as follows. Section \ref{sec:2} presents a new class of residuals and its properties.
In Section \ref{sec:3}, Monte Carlo simulation studies are performed to examine the properties of this class of residuals.
In the following section, an application to real data is shown considering the zero adjusted beta regression model. Concluding remarks are provided in Section \ref{sec:5}. 

\section{A residual for outlier identification}
\label{sec:2}

In this section, we define the zero adjusted regression models and introduce a class of residuals for this class of models.

\subsection{Zero adjusted regression models}
\label{sec:model}

A zero adjusted distributed random variable has probability density function (p.d.f.) given by

\begin{equation}
\label{eq:fdpdt}
f_Y(y;\alpha,\mu,\phi)=
\begin{cases}
\alpha &\text{if} \; y \in \{0\}, \\
(1-\alpha) f_W(y;\mu,\phi) &\text{if} \;y \in \mathcal{Y},
\end{cases}
\end{equation}
where $\alpha = \textrm{Pr}(Y=0)$ is a mixture parameter, $f_W(y;\mu,\phi)$ is the p.d.f. of a continuous distribution with support on $\mathcal{Y} \subset \mathbb{R}^+$, and $\mu$ and $\phi$ are, respectively, location and dispersion (or precision) parameters of $f_W(y;\mu,\phi)$. Using this p.d.f., the class of regression models considered in this paper is defined as following. 


\begin{definition}\label{def:zarm}
Let $Y_1,Y_2,\ldots,Y_n$ be independent random variables, where $Y_i$ is zero adjusted distributed with parameter vector $\theta_i = (\mu_i,\phi_i,\alpha_i)^\top$. 
The zero adjusted regression (ZAR) models are defined by (\ref{eq:fdpdt}) and the following systematic components
\begin{equation}
\label{eq:rinfl}
g_j(\theta_{ij}) = \left. \begin{cases}
g_1(\mu_i) =  \eta_{i1} \\
g_2(\phi_i) = \eta_{i2} \\
g_3(\alpha_i) = \eta_{i3}
\end{cases}\!\!\!\!\!\!\!\!\right\} = \eta_{ij}, \quad i=1,2,\ldots, n; j=1,2,3,
\end{equation}
where $\eta_{ij}= x_{ij}^\top \beta_j$ are linear predictors,  $\beta_j$ is a $p_j \times 1$ vector of unknown parameters, 
$x_{ij}^{\top}=(x_{ij1},x_{ij2},\ldots,x_{ijp_j})$ represents the values of $p_j$ predictor variables and the link functions $g_j(\cdot)$ are strictly monotone and at least twice differentiable. 
\end{definition}

Some particular cases of the model (\ref{eq:rinfl}) are the zero adjusted beta (ZABE) regression model \citep{cook2008regression,Ospina:2012aa}, the zero adjusted gamma (ZAGA) regression model \citep{Tong:2013aa}, the zero adjusted gaussian inverse (ZAIG) regression model \citep{Hinde2006aa} and the zero adjusted reparameterized Birnbaum-Saunders (ZARBS) regression model \citep{tomazella2018zero}. The model (\ref{eq:rinfl}) is also a particular case of the generalized additive models for location, scale and shape \citep{rig+sta05}.
Parameters of the ZAR model can be estimated by maximum likelihood using a numerical nonlinear optimization algorithm \citep[Section 6.1]{noc+wri06}. Confidence intervals regarding the parameters can be obtained using the asymptotic properties of the maximum likelihood estimator and hypothesis testing can be conducted using Wald, likelihood ratio or score statistic \citep[Section 8.6]{sen+sin+lim10}.

\subsection{The randomized quantile residual}
\label{subsec:quantresid}

The randomized quantile residual was introduced by \citet{Dunn:1996aa}. For the ZAR models, it is defined as
	\begin{equation}\label{def:rqr}
	r_i^{q}=\left\{
	\begin{array}{ccc}
	\Phi^{-1}(u_i), & \text{if} & y_i=0, \\	
	\Phi^{-1}(F(y_i;\hat{\theta}_i)), & \text{if} &  y_i>0,
	\end{array}
	\right.
	\end{equation}
where $\hat{\theta}_i$ is the maximum likelihood estimators of $\theta_i$, $\Phi(\cdot)$ is the cumulative distribution function of the standard normal distribution, $F(\cdot)$ is the cumulative distribution function of the response variable and $u_i$ is a uniformly distributed random variable on the interval $(0,\hat{\alpha}_i)$. The residual $r_i^{q}$ is interesting for checking the overall adequacy of a ZAR model when sample size is not small, because it is asymptotically standard normally distributed.

Note from (\ref{def:rqr}) that $r_i^{q} \geq \Phi^{-1}(\hat{\alpha}_i)$ if $y_i > 0$, because, in this case, $F(y_i;\hat{\theta}_i) \geq F(0;\hat{\theta}_i) = \hat{\alpha}_i$. Therefore, if $y_i$ is close to zero and $\hat{\alpha}_i$ is not very small and below 0.5, $r_i^{q}$ will be negative and its absolute value will not be large. For this reason, $r_i^{q}$ may fail to identify some outliers in ZAR models. Suppose for example that a commercial bank want to model the amount of investment that customers have in the bank twelve month after opening a checking account. An important predictor variable in this case is the customer income. If a customer have a very high income but an amount of investment close to zero, this observation should be classified as an outlier. However, using $r_i^{q}$, this observation will not be classified as an outlier if $\hat{\alpha}_i$ is not very small. Rich customers may have $\hat{\alpha}_i$ not so close to zero, because many of them may prefer to invest their money on investment banks.   

\subsection{A residual for outlier identification in ZAR models}
\label{subsec:newresid}

Considering the limitations of $r_i^{q}$, for outlier identification, it is interesting to use a different residual for the discrete and for the continuous component of the model. As mentioned earlier, for the discrete component, a residual defined for regression models for binary dependent variables~\cite[][Chapter 5]{Hosmer_2000} can be used. Similarly, for the continuous component, we can use a residual commonly used in generalized linear models if the fitted model is the ZAGA or the ZAIG regression model or a residual usually used in beta regression if the fitted model is the ZABE regression model. However, for observations that assumed a non-zero value, the residual will not be a function of the estimate of the probability of the observation assuming the zero value and therefore it may not be adequate in some situations. Suppose for example, that we are fitting a ZAGA regression model and we are using the deviance residual for the continuous component of the model. Assume also that, for observations $s$ and $t$, $y_s$ and $y_t$ are positive and $r_t^{d} = r_s^{d} = 3.5$, where $r_i^{d}$ is the deviance residual for observation $i$ considering the continuous component of the ZAGA regression model. In these case, both observations would be considered equally discrepant. Nonetheless, if $\hat{\alpha}_s$ is considerably greater than $\hat{\alpha}_t$ then observation $s$ should be considered more discrepant than observation $t$. 

According to the previous considerations, it seems adequate to define a residual that is a function of $\hat{\alpha}_i$ and $r_i$, where $r_i$ is any residual used in regression models for continuous response variables. The estimate $\hat{\alpha}_i$ is greater than zero and lower than one and $r_i$ can assume any real number. To make them comparable and inspired by the randomized quantile residual, we propose to apply the standard normal distribution function on the absolute value of $r_i$. Therefore, we propose the following residual for outlier identification in ZAR models:
$$
\label{eq:fdpdt2}
r_i^\star=
\begin{cases}
\Phi^{-1}[(1 - \Phi(|r_i|))(1-\hat{\alpha}_i)], &\text{if $r_i < 0$}, \\
\Phi^{-1}[1 - (1 - \Phi(|r_i|))(1-\hat{\alpha}_i)], &\text{if $r_i > 0$}.
\end{cases}
$$
%
The residual $r_i^\star$ is easily rewritten as
\begin{equation}
\label{eq:fdpdt3}
r_i^\star=
\begin{cases}
\Phi^{-1}[\Phi(r_i)(1-\hat{\alpha}_i)], &\text{if $r_i < 0$}, \\
\Phi^{-1}[\hat{\alpha}_i + \Phi(r_i)(1-\hat{\alpha}_i)], &\text{if $r_i > 0$}.
\end{cases}
\end{equation}
Note that (\ref{eq:fdpdt3}) defines a class of residuals for ZAR models, since $r_i$ is any residual used in regression models for continuous response variables (e.g. deviance residual).

A natural choice for $r_i$ is the quantile residual. We will use the term quantile residual and not randomized quantile residual to make clear that it refers to the continuous component of the ZAR model, in which it is not necessary to generate uniformly distributed random variables. We denote $r_i^\star$ as  $r_i^{\star q}$ when $r_i$ is the quantile residual and we named it as zero adjusted quantile residual (ZAQR). For $r_i^{\star q}$, the following theorem holds. The proof is in appendix.

\begin{thm}
\label{teo:quant}
If $r_i^q >0$, then $r_i^{\star q} = r_i^q$.  
\end{thm}

Theorem \ref{teo:quant} suggests that $r_i^{\star q}$ can be interpreted as a correction for outlier identification of the randomized quantile residual when $r_i^q <0$.

Another result about $r_i^\star$ also proved in appendix is stated in the following theorem.

\begin{thm}
\label{teo:norm}
If $r_i$ is standard normally distributed and if $\alpha_i$ is known $\forall i$ then $\forall k > \Phi^{-1}(1 - 0.5(1 - \alpha_i)) = \Phi^{-1}(0.5 + 0.5\alpha_i)$, 
\begin{equation*}
 \textrm{Pr}(r_i^\star < -k) = \textrm{Pr}(r_i^\star > k) = 1-\Phi(k).
 \end{equation*}
 \end{thm}

 According to Theorem  \ref{teo:norm}, if $r_i \sim N(0,1)$ and $\alpha_i$ is known, the residual $r_i^\star$ is quite interesting because it has a behavior similar to a normally distributed random variable. A standard normally distributed residual is desirable because it facilitates the definition of a threshold to the residuals.  However, in general, the distribution of $r_i$ is not normal (not even the quantile residual is standard normally distributed in small samples
\citep{scudilio2017adjusted}) and the parameter $\alpha_i$ is unknown. Thus, to study if $\textrm{Pr}(r_i^\star > k)$ is close to  $1-\Phi(k)$ in practical situations, it is necessary to use Monte Carlo simulation studies.

\section{Simulation studies}
\label{sec:3}

We performed Monte Carlo simulation studies considering the ZABE regression model and using the Ox language \citep{dor09}. For each case, we used 25,000 replications with $n=100$.

In the first scenario, we considered the following ZABE regression model:
$$
\label{eq:rbizutsim}
\begin{cases}
\log\left(\dfrac{\mu_i}{1-\mu_i}\right) &= \beta_{11} + \beta_{12}x_{i12} + \beta_{13}x_{i13}, \\
\log(\phi_i) &= \beta_{21}, \\
\log\left(\dfrac{\alpha_i}{1-\alpha_i}\right) &= \beta_{31} + \beta_{32}x_{i32} + \beta_{33}x_{i33},
\end{cases}
$$
where $\beta_{11} = -1.5$, $\beta_{12} = -1.0$, $\beta_{13} = 1.0$, $\beta_{21} = 4.0$, $\beta_{31} = -0.5$,  $\beta_{32} = 0.5$ and $\beta_{33} = -1.0$, which results in $\mu \in (0.076,0.378)$, $\phi=54.6$ and $\alpha \in (0.182,0.500)$.
 The explanatory variables $x_{i12}$ and $x_{i13}$ were generated as independent draws from the the standard uniform distribution and we assumed $x_{i32}=x_{i12}$ and $x_{i33}=x_{i13}$.  They remained constant throughout the simulations. 
 
In the simulation studies, we compare $r_i^{\star q}$ with four other residuals. The first three belongs to the class introduced in this paper, but using other residuals as $r_i$ instead of the quantile residual. We used the standardized weighted residual 1 \citep{esp+fer+cri08a}, the standardized weighted residual 2 \citep{esp+fer+cri08a} and the adjusted standardized weighted residual 1 \citep{anh+san+bot14} and they are denoted, respectively, by $r_i^{\star efc1}$, $r_i^{\star efc2}$ and $r_i^{\star asb}$. The last one, $r_i^{of}$, is the residual proposed by \citet{Ospina:2012aa} mentioned earlier.

For each of the observations and residuals, we calculated the percentage of the residuals smaller than $-3,-2,-1$ and greater than $1,2,3$ among the $25,000$ replications. When $y_i = 0$, the residuals are not defined and we considered that they are neither smaller than -1 and nor greater than 1. The goal of the simulation studies is to verify if the behavior of the residuals considered in the analysis are similar to the standard normal distribution. For this reason, we calculated descriptive statistics for the $n$ observations of each of the percentages mentioned above and compared them with the theoretical values of the standard normal distribution. 

Table \ref{ta:simZABE1} presents the simulation results for Scenario 1. The mean and median of the calculated percentages for $r_i^{\star q}$ are close to the theoretical values of the standard normal distribution for all intervals. If we consider sampling variability, even the minimum and the maximum of the calculated percentages for $r_i^{\star q}$ are not so far from the theoretical values of the standard normal distribution. On the other hand, none of the other residuals has mean and median of the calculated percentages close to the theoretical values of the standard normal distribution for the 6 considered intervals.

Many other scenarios were evaluated changing the values of the parameters, the distribution of the explanatory variables, sample size and including explanatory variables in the submodel for the parameter $\phi$. Tables were omitted for the sake of brevity. In all scenarios, the distribution of $r_i^{\star q}$ is closer to the standard normal distribution than that of the other residuals. Additionally, in general, the results for $r_i^{\star q}$ are similar to Scenario 1. However, when sample sizes reduces to 50 or when we include explanatory variables in the submodel for the parameter $\phi$, the calculated percentages for $r_i^{\star q}$ are slightly far from the theoretical values of the standard normal distribution than in Scenario 1.

\begin{table}
	\centering
	\caption{Descriptive statistics for the percentage of residuals in each interval - ZABE regression model - Scenario 1}
\scalebox{0.9}{
		\begin{tabular}{ccrrrrrrr}
			\hline
			& \textbf{Residual} & \multicolumn{1}{c}{\textbf{Theoretical}} & \multicolumn{6}{c}{\textbf{Simulation results}} \\
			\cline{4-9}\textbf{Residual} & \textbf{interval} & \multicolumn{1}{c}{\textbf{Value}} & \multicolumn{1}{c}{\textbf{Min}} & \multicolumn{1}{c}{\textbf{Q1}} & \multicolumn{1}{c}{\textbf{Median}} & \multicolumn{1}{c}{\textbf{Mean}} & \multicolumn{1}{c}{\textbf{Q3}} & \multicolumn{1}{c}{\textbf{Max}} \\
			\hline
$r_i^{\star q}$	&	$< -3$	&	0.13	&	0.04	&	0.10	&	0.11	&	0.11	&	0.12	&	0.20	\\
	&	$< -2$	&	2.28	&	1.84	&	2.15	&	2.23	&	2.24	&	2.31	&	2.53	\\
	&	$< -1$	&	15.87	&	15.41	&	15.90	&	16.05	&	16.04	&	16.16	&	16.73	\\
	&	$> 1$	&	15.87	&	15.19	&	15.82	&	15.99	&	16.00	&	16.21	&	16.70	\\
	&	$> 2$	&	2.28	&	1.89	&	2.21	&	2.27	&	2.28	&	2.37	&	2.56	\\
	&	$> 3$	&	0.13	&	0.06	&	0.10	&	0.12	&	0.11	&	0.13	&	0.19	\\
\hline																	
$r_i^{\star asb}$	&	$< -3$	&	0.13	&	0.13	&	0.20	&	0.24	&	0.24	&	0.27	&	0.37	\\  
	&	$< -2$	&	2.28	&	2.30	&	2.55	&	2.62	&	2.63	&	2.73	&	2.96	\\
	&	$< -1$	&	15.87	&	14.80	&	15.33	&	15.59	&	15.58	&	15.80	&	16.35	\\
	&	$> 1$	&	15.87	&	15.33	&	16.03	&	16.28	&	16.27	&	16.44	&	17.09	\\
	&	$> 2$	&	2.28	&	1.48	&	1.75	&	1.81	&	1.84	&	1.93	&	2.26	\\
	&	$> 3$	&	0.13	&	0.00	&	0.03	&	0.04	&	0.04	&	0.06	&	0.12	\\
\hline																	
$r_i^{\star efc1}$	&	$< -3$	&	0.13	&	0.12	&	0.20	&	0.25	&	0.24	&	0.28	&	0.37	\\
	&	$< -2$	&	2.28	&	2.24	&	2.54	&	2.64	&	2.63	&	2.75	&	3.03	\\
	&	$< -1$	&	15.87	&	14.77	&	15.34	&	15.59	&	15.58	&	15.81	&	16.34	\\
	&	$> 1$	&	15.87	&	15.41	&	16.10	&	16.28	&	16.26	&	16.44	&	17.06	\\
	&	$> 2$	&	2.28	&	1.39	&	1.73	&	1.84	&	1.83	&	1.96	&	2.22	\\
	&	$> 3$	&	0.13	&	0.00	&	0.02	&	0.04	&	0.04	&	0.06	&	0.10	\\
\hline																	
$r_i^{\star efc2}$	&	$< -3$	&	0.13	&	0.16	&	0.23	&	0.26	&	0.27	&	0.30	&	0.40	\\
	&	$< -2$	&	2.28	&	2.43	&	2.67	&	2.76	&	2.77	&	2.87	&	3.14	\\
	&	$< -1$	&	15.87	&	14.96	&	15.56	&	15.79	&	15.79	&	16.04	&	16.64	\\
	&	$> 1$	&	15.87	&	15.59	&	16.30	&	16.50	&	16.50	&	16.65	&	17.27	\\
	&	$> 2$	&	2.28	&	1.56	&	1.86	&	1.97	&	1.97	&	2.09	&	2.39	\\
	&	$> 3$	&	0.13	&	0.00	&	0.04	&	0.05	&	0.05	&	0.07	&	0.12	\\
\hline																	
$r_i^{of}$	&	$< -3$	&	0.13	&	0.36	&	0.56	&	0.75	&	0.74	&	0.91	&	1.31	\\
	&	$< -2$	&	2.28	&	3.19	&	3.63	&	3.92	&	3.87	&	4.10	&	4.51	\\
	&	$< -1$	&	15.87	&	11.86	&	13.24	&	13.90	&	13.87	&	14.61	&	15.57	\\
	&	$> 1$	&	15.87	&	12.29	&	13.73	&	14.39	&	14.32	&	14.92	&	15.74	\\
	&	$> 2$	&	2.28	&	2.68	&	2.99	&	3.09	&	3.09	&	3.19	&	3.46	\\
	&	$> 3$	&	0.13	&	0.17	&	0.24	&	0.29	&	0.30	&	0.34	&	0.47	\\
			\hline
		\end{tabular}}
    \label{ta:simZABE1}		
\end{table}			

We also performed Monte Carlo simulation studies considering the ZAIG and the ZAGA regression model. Here, we considered as $r_i$ the quantile residual and four others: deviance residual \citep{davison}, the Pearson residual \citep{McCullagh:1989aa}, the Williams residual \citep{williams} and the Anscombe residual \citep{pierce1986residuals}. They are denoted as \adev, \apea, \awil and \aasc, respectively. We compared these residuals with \rqu, a natural extension to ZAGA and ZAIG regression model of the residual proposed by \citet{Ospina:2012aa} for the ZABE regression model, but here using the quantile residual.

Many scenarios were also considered, but for the sake of brevity we only present the results of Scenario 1 for the ZAIG regression model (Table \ref{ta:simZAIG1}). In this scenario, the explanatory variables were generated as independent draws from the the standard uniform distribution and the values of the parameters result in $\mu \in (20.9,403.4)$, $\phi=0.02$ and $\alpha \in (0.27,0.62)$. As in the ZABE regression model, the calculated percentages for $r_i^{\star q}$ are close to the theoretical values of the standard normal distribution for all intervals. On the other hand, the other residuals has calculated percentages far from standard normal values for some intervals. In the other scenarios and in the scenarios for the ZAGA regression model, the distribution of $r_i^{\star q}$ is also closer to the standard normal distribution than that of the other residuals. In addition, in all of the scenarios for ZAIG or ZAGA regression model, the calculated percentages are not far from the theoretical values of the standard normal distribution.

\begin{table}
	\centering
	\caption{Descriptive statistics for the percentage of residuals in each interval - ZAIG regression model - Scenario 1}
\scalebox{0.9}{
		\begin{tabular}{ccrrrrrrr}
			\hline
			& \textbf{Residual} & \multicolumn{1}{c}{\textbf{Theoretical}} & \multicolumn{6}{c}{\textbf{Simulation results}} \\
			\cline{4-9}\textbf{Residual} & \textbf{interval} & \multicolumn{1}{c}{\textbf{Value}} & \multicolumn{1}{c}{\textbf{Min}} & \multicolumn{1}{c}{\textbf{Q1}} & \multicolumn{1}{c}{\textbf{Median}} & \multicolumn{1}{c}{\textbf{Mean}} & \multicolumn{1}{c}{\textbf{Q3}} & \multicolumn{1}{c}{\textbf{Max}} \\
			\hline
			& $< -3$   & 0.13  & 0.06  & 0.10  & 0.12  & 0.12  & 0.13  & 0.18 \\
			& $< -2$   & 2.28  & 1.54  & 2.22  & 2.33  & 2.30  & 2.41  & 2.64 \\
			$r_i^{\star q}$  & $< -1$   & 15.87 & 14.99 & 15.76 & 16.00 & 15.94 & 16.17 & 16.62 \\
			& $> 1$    & 15.87 & 14.46 & 15.84 & 16.03 & 16.01 & 16.20 & 16.86 \\
			& $> 2$    & 2.28  & 1.58  & 2.16  & 2.26  & 2.23  & 2.36  & 2.56 \\
			& $> 3$    & 0.13  & 0.03  & 0.08  & 0.10  & 0.09  & 0.11  & 0.16 \\			
			\hline
			& $< -3$   & 0.13  & 0.11  & 0.16  & 0.18  & 0.18  & 0.21  & 0.28 \\
			& $< -2$   & 2.28  & 2.24  & 2.78  & 2.92  & 2.92  & 3.09  & 3.54 \\
			\adev & $< -1$   & 15.87 & 16.09 & 17.28 & 17.79 & 17.87 & 18.51 & 20.33 \\
			& $> 1$    & 15.87 & 12.21 & 13.98 & 14.54 & 14.34 & 14.76 & 15.71 \\
			& $> 2$    & 2.28  & 1.35  & 1.78  & 1.88  & 1.87  & 1.97  & 2.14 \\
			& $> 3$    & 0.13  & 0.03  & 0.06  & 0.07  & 0.07  & 0.08  & 0.12 \\
			\hline
			& $< -3$   & 0.13  & 0.00  & 0.00  & 0.00  & 0.01  & 0.01  & 0.04 \\
			& $< -2$   & 2.28  & 0.09  & 0.92  & 1.40  & 1.24  & 1.57  & 1.91 \\
			\apea & $< -1$   & 15.87 & 15.98 & 16.84 & 17.09 & 17.04 & 17.32 & 17.83 \\
			& $> 1$    & 15.87 & 14.08 & 14.71 & 14.95 & 14.93 & 15.14 & 15.98 \\
			& $> 2$    & 2.28  & 2.20  & 2.90  & 3.11  & 3.10  & 3.30  & 3.92 \\
			& $> 3$    & 0.13  & 0.11  & 0.34  & 0.44  & 0.47  & 0.58  & 0.96 \\
			\hline
			& $< -3$   & 0.13  & 0.06  & 0.12  & 0.13  & 0.13  & 0.15  & 0.20 \\
			& $< -2$   & 2.28  & 1.74  & 2.60  & 2.75  & 2.72  & 2.88  & 3.16 \\
			\aasc & $< -1$   & 15.87 & 15.80 & 17.12 & 17.69 & 17.70 & 18.39 & 20.06 \\
			& $> 1$    & 15.87 & 11.94 & 13.79 & 14.36 & 14.20 & 14.63 & 15.54 \\
			& $> 2$    & 2.28  & 1.18  & 1.60  & 1.74  & 1.72  & 1.85  & 2.02 \\
			& $> 3$    & 0.13  & 0.01  & 0.04  & 0.06  & 0.06  & 0.07  & 0.10 \\
			\hline
			& $< -3$   & 0.13  & 0.10  & 0.15  & 0.17  & 0.17  & 0.20  & 0.26 \\
			& $< -2$   & 2.28  & 2.18  & 2.72  & 2.89  & 2.88  & 3.06  & 3.46 \\
			\awil & $< -1$   & 15.87 & 16.09 & 17.27 & 17.78 & 17.85 & 18.50 & 20.27 \\
			& $> 1$    & 15.87 & 12.27 & 14.02 & 14.55 & 14.36 & 14.77 & 15.72 \\
			& $> 2$    & 2.28  & 1.39  & 1.82  & 1.91  & 1.91  & 2.01  & 2.19 \\
			& $> 3$    & 0.13  & 0.04  & 0.07  & 0.08  & 0.08  & 0.09  & 0.13 \\
			\hline
			& $< -3$   & 0.13  & 0.44  & 0.60  & 0.72  & 0.71  & 0.81  & 0.96 \\
			& $< -2$   & 2.28  & 3.25  & 3.66  & 3.81  & 3.78  & 3.93  & 4.20 \\
			\rqu  & $< -1$   & 15.87 & 9.86  & 12.09 & 12.84 & 12.69 & 13.68 & 14.76 \\
			& $> 1$    & 15.87 & 9.66  & 12.16 & 12.92 & 12.73 & 13.42 & 14.57 \\
			& $> 2$    & 2.28  & 3.10  & 3.55  & 3.74  & 3.72  & 3.87  & 4.21 \\
			& $> 3$    & 0.13  & 0.32  & 0.54  & 0.65  & 0.65  & 0.75  & 0.96 \\
			\hline
					\end{tabular}}
    \label{ta:simZAIG1}		
\end{table}
    
\section{Application}
\label{sec:4}

In this section we use data from 1000 applicants of the National Exam of Upper Secondary Education (ENEM) from 2014. The brazilian exam was created in 1998 as an alternative to higher education entrance examinations and focus on a comprehensive and structural concept of human intelligence. Data were obtained from \url{http://inep.gov.br/web/guest/microdados}. 

The exam has 5 different tests: natural sciences, human sciences, language, mathematics and writing essay. The score of an applicant on the writing essay test is limited on the interval $[0;1000]$. Here we considered the score on the writing essay test divided by 1000 and our goal is to study the relation of this response variable and the scores of the other tests, the gender and the age of the applicant.  In our sample, none of the applicants reached the maximum score and 51 obtained a zero score. Based on this information, the ZABE regression model is a natural choice to fit the response variable. Considering the results of exploratory analysis, the age of the applicant was categorized in two levels: greater than 25 and others.

\begin{figure}[!htbp]\centering
\includegraphics[width=10cm,height=10cm]{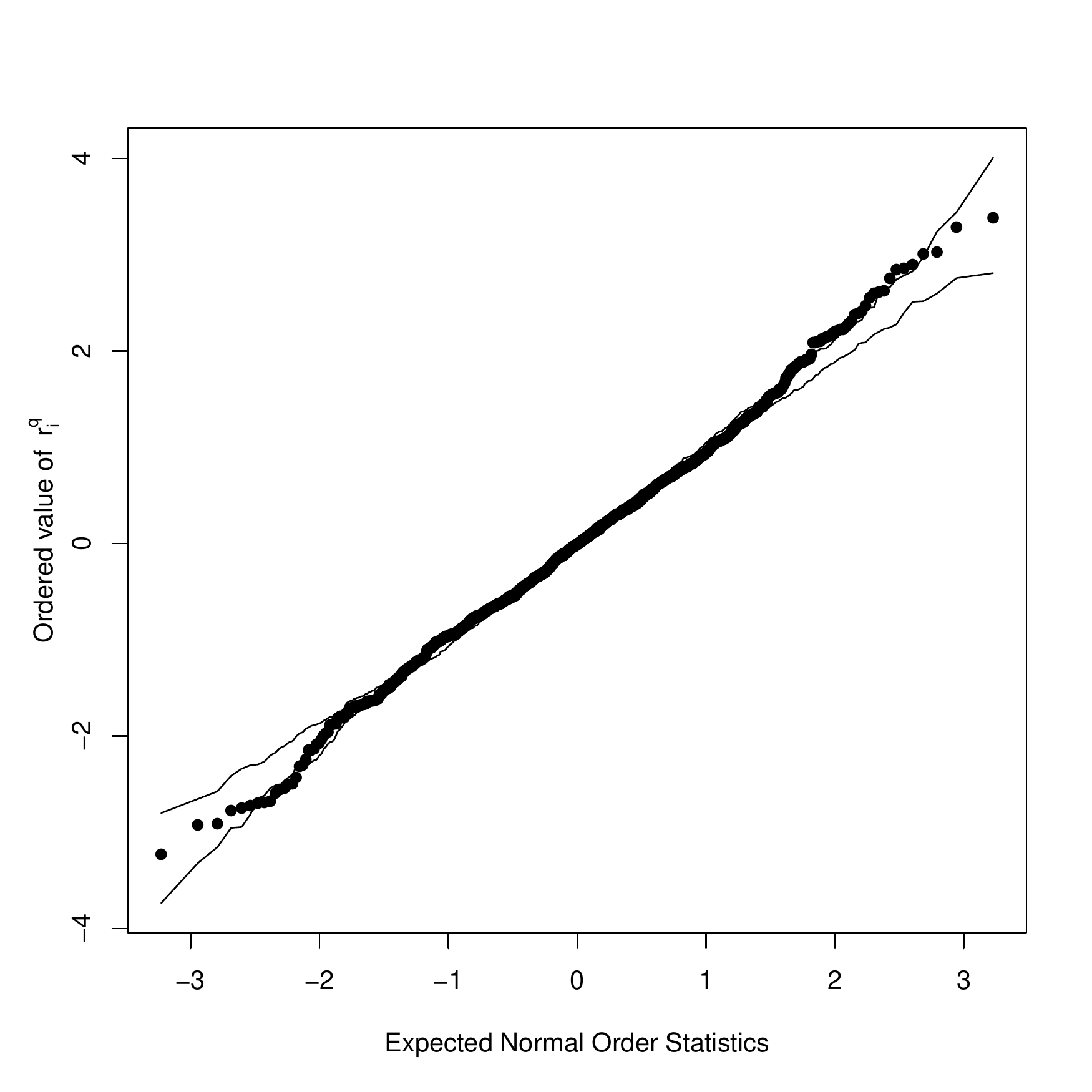}
\caption{Normal probability plot with simulated envelope for the randomized quantile residual.}
\label{fi:normplot}
\end{figure}

All variables were initially included in the three submodels of the ZABE regression model. Variables were selected for the final model in a backward way using Wald tests. Figure \ref{fi:normplot} presents a half-normal residual plot with simulated envelope \citep[Section 4.2]{atk85} for the final model using $r_i^{q}$. The plot does not suggest model misspecification.

\begin{figure}[!htbp]\centering
\includegraphics[width=17cm,height=8.5cm]{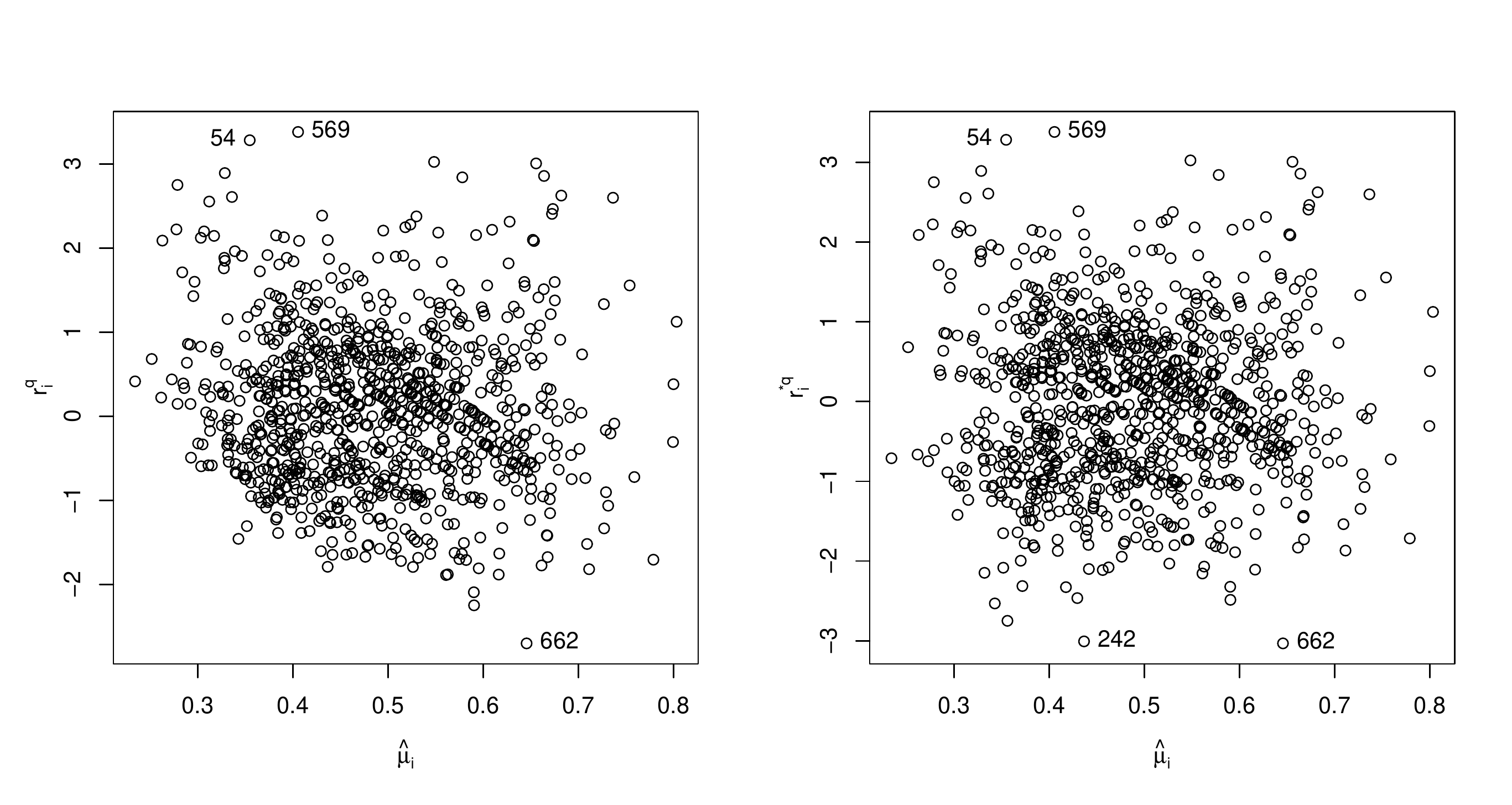}
\caption{Residuals against $\hat{\mu}_i$ for $r^q_i$ (left) and $r_i^{\star q}$ (right).}
\label{fi:residplot}
\end{figure}

To perform outlier identification, Figure \ref{fi:residplot} presents plots of residuals against $\mu_i$ considering $r_i^{q}$ (left) and $r_i^{\star q}$ (right) and using the 949 applicants that obtained a score in the writing essay test greater than zero. None of the observations has absolute value of the residuals considerable greater than the others. However, using $r_i^{\star q}$, four observations stand out. Three of them also emerge when $r_i^{q}$ is used, but observation \#242 does not have a high absolute value of $r_i^{q}$. This applicant obtained score 0.12 on the writing essay test, considerably lower than the fitted value $\hat{\mu}_{242}=0.437$. However, as $\hat{\alpha}_{242}$ is not very low $(0.037)$, the absolute value of $r_i^{q}$ is not high for this applicant and this residual fails to identify one of the worst fitted observations. 

We fitted the final ZABE regression model excluding, one at a time, the four observations pointed out in Figure \ref{fi:residplot} and compared the estimates of the parameters of each of these models with that obtained using all observations. The changes in the submodel for the parameter $\mu$ are lower than $6\%$. In addition, only when we exclude observation \#662, the change in one of the estimates in the submodel for the parameter $\phi$ is greater than $8\%$, and it is not so high $(13.3\%)$. As we considered in Figure \ref{fi:residplot} only the 949 applicants that obtained a score in the writing essay test greater than zero, the changes in the estimates in the submodel for the parameter $\alpha$ are very small. We also analyzed the observation with zero value in the response variable and none of these observations have absolute value of the deviance residual greater than 3. Therefore, there are not observations that affect considerably the results of the final ZABE regression model.

Assume now that the value of the response variable for observation \#242 was incorrectly recorded as 0.001, the minimum possible score greater than zero. In this case, after fitting the ZABE regression model, we obtain $r_i^{q}=-1.81$ and $r_i^{\star q}=-7.86$. Note that $r_i^{\star q}$ clearly identify the observation as  outlier, while $r_i^{q}$ fails to do it. Considering the ZABE regression model, the estimate of the probability of an observation  assuming a value greater than zero and lower than 0.01 if it has the covariates of the observation \#242 is very small $(1.7 \times 10^{-9})$. Excluding this observation, four estimates of parameters change more than $14\%$, two of them are in submodel for $\mu$, which is the main one in a ZABE regression model. In addition, the change in the estimate of the parameter associated with human sciences score in the submodel for $\phi$ is very high $(52\%)$. Based on these information, observation \#242 should be identified as an outlier, which could enable to an applied statistician identifying that this observations was incorrectly recorded. This example illustrates that $r_i^{\star q}$ is better than  $r_i^{q}$ for outlier identification in zero adjusted regression models.

Table \ref{ta:aplic} presents the results of the final ZABE regression model for the writing essay score. As expected, the mean of the continuous component of the distribution of the writing essay score increases as the score in the other tests increase. However, only for two out of the four other tests, there are evidences that an increase in the score of these tests, reduces the probability of the applicant obtaining a zero score in the writing essay test. In addition, older applicants has a lower mean and a higher probability of obtaining a zero score, while female applicants has a better mean performance.

     \begin{table} [!hpb]
    \caption {The final ZABE regression model for the writing essay score}
    \begin{center}
 \begin{tabular} {ccrrr}
\hline									
Equation	&	Variable	&	Estimate	&	Standard	&	P-value	\\
	&		&		&	error	&		\\
\hline									
$\mu$	&	Intercept	&	-3.3924	&	0.1511	&	$< 0.0001$	\\
	&	Natural sciences score	&	0.0015	&	0.0004	&	$< 0.0001$	\\  
	&	Human sciences score	&	0.0019	&	0.0004	&	$< 0.0001$	\\
	&	Language score	&	0.0025	&	0.0004	&	0.0011	\\
	&	Mathematics score	&	0.0009	&	0.0003	&	$< 0.0001$	\\
	&	Male student	&	-0.1968	&	0.0351	&	0.0004	\\
	&	Age greather than 25	&	-0.1417	&	0.0396	&	$< 0.0001$	\\
\hline									
$\phi$	&	Intercept	&	-1.4090	&	0.1751	&	$< 0.0001$	\\
	&	Human sciences score	&	-0.0009	&	0.0003	&	0.0084	\\
	&	Language score	&	0.0016	&	0.0004	&	$< 0.0001$	\\
\hline									
$\alpha$	&	Intercept	&	6.0367	&	1.1238	&	$< 0.0001$	\\
	&	Human sciences score	&	-0.0094	&	0.0025	&	0.0002	\\
	&	Language score	&	-0.0094	&	0.0028	&	0.0006	\\
	&	Age greater than 25	&	0.7444	&	0.3166	&	0.0189	\\
\hline									
    \end{tabular}
    \end{center}
    \label{ta:aplic}
    \end{table}

\section{Concluding remarks}
\label{sec:5}

In this work, we introduced a class of residuals for zero adjusted regression models. Monte Carlo simulation studies and an application were performed to compare these class of residuals with other residuals that can be used to perform diagnostic analysis in these regression models. 
The simulation studies suggested that one of the residual of this new class, named ZAQR, has some similar properties to a standard normally distributed variable. 
In addition, the application indicated that ZAQR can identify outliers in situations in which randomized quantile residual fails to do it. For these reasons, ZAQR seems to be the best option for outlier identification in zero adjusted regression models.

It is also interesting to mention that the randomized quantile residual may also fail to identify some outliers in regression models in which the response variable is discrete in more than one point, as the zero-and-one adjusted beta regression model \citep{Ospina08,stasinopoulos2007generalized} and the three-point adjusted beta regression model \citep{pereira2013regression,gray2018command}. Therefore, future works can be developed extending ZAQR for outlier identification in these class of models.


\section*{Acknowledgments}
This study was financed in part by the Coordenação de Aperfeiçoamento de Pessoal de Nível Superior - Brazil (CAPES) - Finance Code 001 and by São Paulo Research Foundation (FAPESP) - Grant \#2013/17876-6. 

\bibliographystyle{authordate1}

\appendix

\section{Appendix: Proof of theorems of Section \ref{subsec:newresid}}
\label{sec:append}

In this appendix, we prove the theorems presented in Section \ref{subsec:newresid}.

\subsection{Proof of Theorem \ref{teo:quant}}

If $r_i^q >0$ then

\begin{eqnarray*}
r_i^{\star q}&=&\Phi^{-1}\left[\hat{\alpha}_i+ \Phi\left[\Phi^{-1}(\widehat{\textrm{Pr}}(Y_i\leq{y}_i |Y_i>0) )\right] (1-\hat{\alpha}_i)\right]\\
&=& \Phi^{-1}\left[\hat{\alpha}_i+ \widehat{\textrm{Pr}}(Y_i\leq{y}_i|Y_i>0) (1-\hat{\alpha}_i)\right]\\
&=& \Phi^{-1}\left[\hat{\alpha}_i+ \frac{\widehat{\textrm{Pr}}(0<Y_i\leq y_i)}{\widehat{\textrm{Pr}}(y_i>0)} (1-\hat{\alpha}_i)\right]\\
&=& \Phi^{-1}\left[\hat{\alpha}_i+ \frac{\widehat{\textrm{Pr}}(0<Y_i\leq y_i)}{(1-\hat{\alpha}_i)} (1-\hat{\alpha}_i)\right]\\
&=& \Phi^{-1}\left[\widehat{\textrm{Pr}}(Y_i=0) + \widehat{\textrm{Pr}}(0<Y_i\leq y_i) \right]\\
&=& \Phi^{-1}\left[ \widehat{\textrm{Pr}}(Y_i\leq y_i)\right]\\
&=&\Phi^{-1}\left[F(y_i, \hat \theta)\right]= r^q_i.
\end{eqnarray*}

\subsection{Proof of Theorem \ref{teo:norm}}

If $r_i$ is normally distributed with mean 0 and variance 1 and if $\alpha_i$ is known $\forall i$ then $\forall k > \Phi^{-1}(1 - 0.5(1 - \alpha_i)) = \Phi^{-1}(0.5 + 0.5\alpha_i)$, then

 \begin{eqnarray*}
\textrm{Pr}(r_i^*<-k) &=& \textrm{Pr}(\Phi(r_i)(1 - \alpha_i) < \Phi(-k)) \\
                                  &=& \textrm{Pr}\left[r_i < \Phi^{-1}\left(\frac{\Phi(-k)}{(1 -\alpha_i)}\right)\right]\\
                                  &=& \textrm{Pr}\left[r_i < \Phi^{-1}\left(\frac{\Phi(-k)}{(1 -\alpha_i)}\right),y_i > 0\right]\\                                 
                                  &=&\Phi\left[ \Phi^{-1}\left(\frac{\Phi(-k)}{(1 -\alpha_i)}\right)\right](1-\alpha_i)\\
                                  &=&\left[\frac{\Phi(-k)}{(1 -\alpha_i)} (1-\alpha_i)\right]\\
                                  &=&  1-\Phi(k).
 		\end{eqnarray*}

In a similar way, it is shown that $\textrm{Pr}(r_i^\star > k) = 1-\Phi(k)$.

\end{document}